\documentclass[draft]{agujournal2019}
\usepackage{url} 
\usepackage{soul}
\usepackage{amsfonts}
\usepackage{amsmath}
\usepackage{amssymb}
\draftfalse

\journalname{Radio Science}
\newcommand{\Langle}{\left\langle}
\newcommand{\Rangle}{\right\rangle}
\newcommand{\eq}[1]{Eq.~(\ref{#1})}
\newcommand{\eqs}[2]{Eqs.~(\ref{#1}) and~(\ref{#2})}
\newcommand{\fig}[1]{Fig.~\ref{#1}}

\newcommand{\Circ}[1]{\operatorname{circ}\left({#1}\right)}

\begin{document}

%
%


\title{Anisoplanatic errors in two-wavelength adaptive optics}

%
%




\authors{M. W. Hyde\affil{1}, M. F. Spencer\affil{2}, and M. Kalensky\affil{3}}

 \affiliation{1}{Epsilon C5I, Beavercreek, OH 45431 USA}
 \affiliation{2}{The University of Arizona, James C. Wyant College of Optical Sciences, Tucson, AZ 85721 USA}
 \affiliation{3}{Naval Surface Warfare Center Dahlgren
Division, Dahlgren, VA 22448 USA.}





\correspondingauthor{M. Hyde}{mhyde@epsilonsystems.com}



\begin{keypoints}
\item Quantifies the effects of anisoplanatism on two-wavelength adaptive optics systems for imaging and beam projection applications
\item Introduces the two-wavelength isoplanatic angle as a key predictor of system performance 
\item Closed-form expressions for anisoplanatic errors are in excellent agreement with numerical results
\end{keypoints}

%
%

%
%


\begin{abstract}
Two-wavelength adaptive optics (AO) systems sense wavefront errors using a beacon at one wavelength, while correcting for subsequent imaging or beam projection at another.  Although most AO systems operate in this manner, the relevant AO literature is generally concerned with quantifying system performance at a single wavelength, effectively ignoring the two-wavelength nature of the problem.  In this paper, we study the effects of anisoplanatism, or the physical separation of the beacon and transmit light, on two-wavelength AO systems for imaging and beam projection applications.  We derive the piston-removed (PR) and piston-and-tilt-removed (PTR) optical-path-difference (OPD) variances including anisoplanatism, which are key metrics of AO system performance.  We compare our closed-form PR and PTR OPD variances to numerical results and discuss their physical significance.  In addition, we introduce the two-wavelength isoplanatic angle and show how it can be used to quickly assess two-wavelength AO system performance.  At large, the analysis contained in this paper will inform the design and implementation of two-wavelength AO systems for multiple imaging and beam projection applications.            
\end{abstract}


%
%

%


%
%
%
%

\section{Introduction}
Two-wavelength adaptive optics (AO) is when wavefront sensing, using light from an artificial or natural beacon, is performed at one wavelength $\lambda_\text{B}$ and correction (for imaging or beam projection) is performed at another $\lambda_\text{T}$.  This scenario is common in astronomy, where light from laser guide stars excites sodium atoms in the mesosphere forming multiple beacons~\cite{gilmozzi2007european,Sanders2013,Hardy}.  Here, 589.2~nm light is used to sense and correct turbulence-induced wavefront aberrations.  Scientific analysis or imaging, however, is typically performed at longer wavelengths in the short-wave infrared.         

In the late 1970s and early 1980s, contemporary with the development of laser guide stars~\cite{LGSA,Fugate:94,fugate1991measurement,Fugate:23,Parenti:94}, researchers such as \citeA{Lukin:79}, \citeA{Hogge:82}, \citeA{Holmes:83}, \citeA{Winocur:83}, and \citeA{Wallner:84} quantified the residual wavefront errors or drop in Strehl ratio inherent in two-wavelength AO systems.  \citeA{Lukin:79} and \citeA{Hogge:82} derived integral expressions for the two-wavelength phase and optical-path-difference (OPD) variances in weak atmospheric turbulence, respectively.  \citeA{Holmes:83} numerically investigated the reduction in Strehl ratio due to differences in the beacon and transmit wavelengths.  \citeA{Winocur:83} investigated the effects of $\lambda_\text{B}$-$\lambda_\text{T}$ separation on the first five Zernike polynomials/aberrations, with special emphasis on Zernike tilt.  Lastly, \citeA{Wallner:84} compared diffractive OPD error---originally studied by \citeA{Lukin:79} and \citeA{Hogge:82}---to refractive error caused by atmospheric dispersion. 

From this early analysis, a consensus emerged: As long as $\lambda_\text{B} < \lambda_\text{T}$ and the $\lambda_\text{B}$-$\lambda_\text{T}$ separation was less than a few microns, diffractive error was generally negligible for astronomical/vertical viewing.  AO research for astronomy naturally shifted toward quantifying and correcting refractive and anisoplanatic errors,
which are prevalent in wide-field-of-view imaging systems~\cite{Johnston:94,Devaney:08,Wang:12}.  

Recently, there has been interest in applying two-wavelength AO for imaging and beam projection systems that operate in complementary conditions to astronomy, i.e., over small fields of view and long horizontal paths~\cite{photonics10070725,10.4236/opj.2020.104006,Ke2025,Hyde:2024,10529268,2025arXiv250500609H,doi:10.2514/1.J061766,Perram,Merritt}.  As such, design considerations for such systems commonly result in $\lambda_\text{B} > \lambda_\text{T}$, violating one of the conditions derived by early two-wavelength AO researchers.  

Three recent works by us explored wavefront errors in two-wavelength AO systems for imaging and beam projection applications: The first investigated branch point/branch cut~\cite{Fried:92,Fried:98} effects for two-wavelength AO systems operating in strong turbulence~\cite{Hyde:2024}.  The second derived closed-form scaling law formulas for the plane-wave, two-wavelength, piston-removed (PR) OPD variance~\cite{10529268}.  Lastly, the third derived spherical-wave, two-wavelength OPD variance expressions for tilt and higher-order wavefront errors~\cite{2025arXiv250500609H}.    

Although those three works quantified a majority of the relevant errors in two-wavelength AO, they did not consider anisoplanatism (i.e., the physical separation of $\lambda_\text{B}$ and $\lambda_\text{T}$), which is a major source of error in AO~\cite{Fried_1979,Sasiela:07,Beck:20,Bos:24,Kalensky:24}.  To our knowledge, two-wavelength anisoplanatic error has never been quantified before and, as we show in subsequent sections, is a significant theoretical undertaking, well beyond that of our prior works.
\begin{figure}[t]
    \centering
    \includegraphics[width=\linewidth]{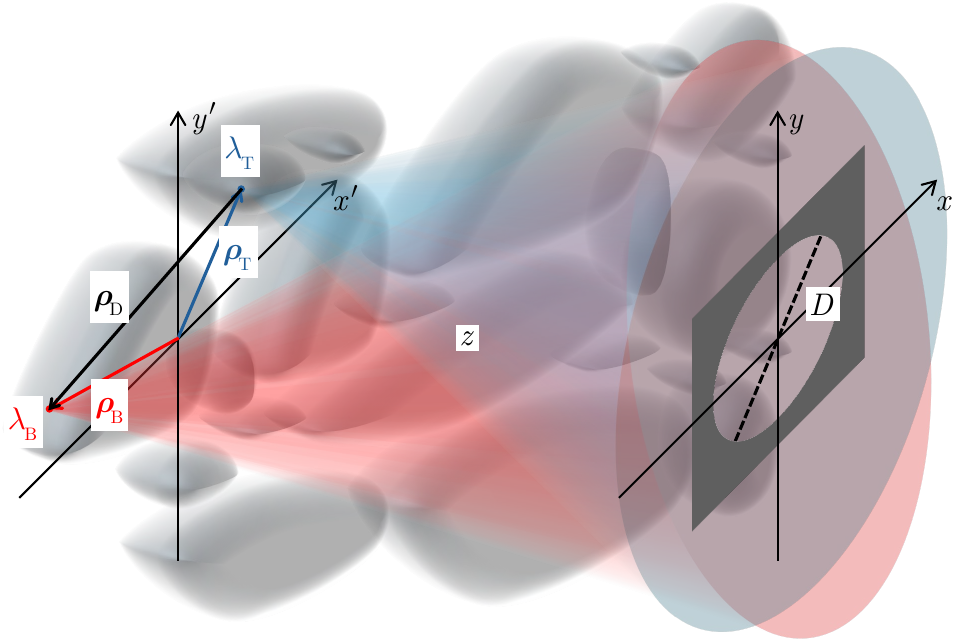}
    \caption{Two-wavelength, anisoplanatic propagation geometry.  Two off-axis point sources---one at wavelength $\lambda_\text{B}$ and location $\boldsymbol{\rho}_\text{B}$ and the other at wavelength $\lambda_\text{T}$ and location $\boldsymbol{\rho}_\text{T}$---emit spherical waves that propagate a distance $z$ through atmospheric turbulence and are received through a circular aperture of diameter $D$.}
    \label{fig:geom}
\end{figure}

In this paper, we quantify the two-wavelength, anisoplanatic PR and piston-and-tilt-removed (PTR) OPD variances, which are applicable to the tilt and higher-order correcting subsystems of two-wavelength AO systems.  The geometry for our subsequent analysis is shown in Figure~\ref{fig:geom}.  The figure depicts two point sources, of differing wavelengths ($\lambda_\text{B}$ and $\lambda_\text{T}$) and off-axis locations ($\boldsymbol{\rho}_\text{B}$ and $\boldsymbol{\rho}_\text{T}$), radiating spherical waves a distance $z$ through atmospheric turbulence.  These waves enter a circular aperture of diameter $D$.  Our goal is to find an expression for the variance of the OPD $\Delta \ell$---or phase difference via
\begin{equation}\label{eq:PR1}
 \phi\left(\lambda\right) = k  \Delta \ell,   
\end{equation}
where $k = 2\pi/\lambda$---between these two point sources as a function of their wavelength and location deltas, i.e., $\lambda_\text{B} - \lambda_\text{T}$ and $\boldsymbol{\rho}_\text{B}-\boldsymbol{\rho}_\text{T}=\boldsymbol{\rho}_\text{D}$.  

In what follows, we begin with the two-wavelength, anisoplanatic PR OPD variance and conclude with the associated PTR OPD variance.  Throughout our analysis, we compare our closed-form expressions to numerical results and discuss their physical significance.  In addition, we introduce the two-wavelength isoplanatic angle---a generalized version of the traditional isoplanatic angle $\theta_0$~\cite{Fried_1979,Sasiela:07}---and discuss how it can be used to gauge two-wavelength AO system performance.  Lastly, we conclude with a summary of our work and contributions.

\section{Anisoplanatic, Two-Wavelength PR OPD Variance}
We begin our analysis with an expression for the PR optical path length $\ell$ over the circular aperture of diameter $D$ (see \fig{fig:geom}):
\begin{equation}\label{eq:PR2}
\ell_\text{PR}\left(\boldsymbol{\rho},z;\lambda,\boldsymbol{\rho}_c\right) = \ell\left(\boldsymbol{\rho},z;\lambda,\boldsymbol{\rho}_c\right) - \frac{1}{A}\iint_{-\infty}^{\infty}\Circ{\frac{\boldsymbol{\rho}}{D/2}}\ell\left(\boldsymbol{\rho},z;\lambda,\boldsymbol{\rho}_c\right) \text{d}^2\rho,
\end{equation}
where $A = \pi\left(D/2\right)^2$ and $\boldsymbol{\rho}_c$ is the off-axis location of the point source.  The PR OPD variance between the two point sources in Figure~\ref{fig:geom}, averaged over the aperture, is
\begin{equation}\label{eq:PR3}
\Langle \Delta \ell_\text{PR}^2 \Rangle = \frac{1}{A}\iint_{-\infty}^{\infty}\Circ{\frac{\boldsymbol{\rho}}{D/2}}\Langle \left[ \ell_\text{PR}\left(\boldsymbol{\rho},z;\lambda_\text{B},\boldsymbol{\rho}_\text{B}\right) -\ell_\text{PR}\left(\boldsymbol{\rho},z;\lambda_\text{T},\boldsymbol{\rho}_\text{T}\right)\right]^2\Rangle \text{d}^2\rho.
\end{equation}
Substituting \eq{eq:PR2} into \eq{eq:PR3}, expanding the squared quantity, and assuming statistically homogeneous turbulence reveals
\begin{equation}\label{eq:PR4}
\begin{gathered}
\Langle \Delta \ell_\text{PR}^2 \Rangle = \frac{1}{2A} \iint_{-\infty}^{\infty} \Lambda\left(\frac{\boldsymbol{\rho}}{D}\right) \left[D_\ell\left(\boldsymbol{\rho},z;\lambda_\text{B},\boldsymbol{\rho}_\text{B};\lambda_\text{B},\boldsymbol{\rho}_\text{B}\right) + D_\ell\left(\boldsymbol{\rho},z;\lambda_\text{T},\boldsymbol{\rho}_\text{T}\lambda_\text{T},\boldsymbol{\rho}_\text{T}\right)\right. \hfill \\
\left. \quad -\, 2D_\ell\left(\boldsymbol{\rho},z;\lambda_\text{B},\boldsymbol{\rho}_\text{B};\lambda_\text{T},\boldsymbol{\rho}_\text{T}\right) \right] \text{d}^2\rho. \hfill
\end{gathered}
\end{equation}
In the above equation, $\Lambda\left(x\right)$ is the optical transfer function~\cite{IFO,Gaskill1978}
\begin{equation}\label{eq:PR5}
\Lambda\left(x\right) = \frac{2}{\pi}\left[\cos^{-1}\left(x\right)-x\sqrt{1-x^2}\right]\operatorname{circ}\left({x}\right), 
\end{equation}
$D_\ell$ is the anisoplanatic, two-wavelength optical-path-length (OPL) structure function
\begin{equation}\label{eq:PR6}
D_\ell\left(\boldsymbol{\rho},z;\lambda_\text{B},\boldsymbol{\rho}_\text{B};\lambda_\text{T},\boldsymbol{\rho}_\text{T}\right) = \frac{2}{k_\text{B}k_\text{T}}\left[ B_S\left(0,z;\lambda_\text{B},\boldsymbol{\rho}_\text{B};\lambda_\text{T},\boldsymbol{\rho}_\text{T}\right)- B_S\left(\boldsymbol{\rho},z;\lambda_\text{B},\boldsymbol{\rho}_\text{B};\lambda_\text{T},\boldsymbol{\rho}_\text{T}\right) \right],
\end{equation}
and $B_S$ is the phase covariance function derived in the Supporting Information~\cite{si1}.  Substituting \eq{eq:PR6}---in combination with $B_S$ from \citeA{si1}---into \eq{eq:PR4} and rearranging the integrals yields
\begin{equation}\label{eq:PR7}
\begin{gathered}
\Langle \Delta \ell_\text{PR}^2 \Rangle = 2\pi \int_0^z \iint_{-\infty}^{\infty} \Phi_n\left(\boldsymbol{\kappa},z'\right) \operatorname{Re}\left(\left\{ \sum\limits_{k=1}^3 c_k \exp\left[-\text{j}\alpha_k \frac{z'}{z}\left(1-\frac{z'}{z}\right) \kappa^2\right] \right. \right. \hfill \\
\left.\left. \quad + \exp\left[\text{j}\left(1-\frac{z'}{z}\right)\boldsymbol{\rho}_\text{D}\cdot\boldsymbol{\kappa}\right] \sum\limits_{k=4}^5 c_k \exp\left[-\text{j}\alpha_k \frac{z'}{z}\left(1-\frac{z'}{z}\right) \kappa^2\right]\right\}  \right. \hfill \\
\left. \quad\, \times \frac{1}{A} \iint_{-\infty}^{\infty} \Lambda\left(\frac{\boldsymbol{\rho}}{D}\right) \left[1-\exp\left(\text{j}\frac{z'}{z}\boldsymbol{\kappa}\cdot\boldsymbol{\rho}\right)\right]\text{d}^2\rho \right)\text{d}^2\kappa \text{d}z', \hfill   
\end{gathered}    
\end{equation}
where $\boldsymbol{\rho}_\text{D}=\boldsymbol{\rho}_\text{B}-\boldsymbol{\rho}_\text{T}$, $\Phi_n$ is the index of refraction power spectrum, $\boldsymbol{c} = \begin{bmatrix}1,1/2,1/2,-1,-1\end{bmatrix}$, and 
\begin{equation}
\boldsymbol{\alpha} = \begin{bmatrix} 0,\dfrac{z}{k_\text{B}},\dfrac{z}{k_\text{T}},\dfrac{z}{2}\left(\dfrac{1}{k_\text{B}}-\dfrac{1}{k_\text{T}}\right),\dfrac{z}{2}\left(\dfrac{1}{k_\text{B}}+\dfrac{1}{k_\text{T}}\right) \end{bmatrix}.   
\end{equation}
The integrals over $\boldsymbol{\rho}$ in \eq{eq:PR7} equal $1-\operatorname{jinc}^2\left[D z' \kappa/\left(2z\right)\right]$, where $\operatorname{jinc}\left(x\right) = 2 J_1\left(x\right)/x$.  Finally, assuming isotropic turbulence, we arrive at the general integral expression for the anisoplanatic, two-wavelength PR OPD variance: 
\begin{equation}\label{eq:PR9}
\begin{gathered}
\Langle \Delta \ell_\text{PR}^2 \Rangle = 4\pi^2 \sum\limits_{k=1}^3 c_k \int_0^z\int_0^\infty \kappa \Phi_n\left(\kappa,z'\right) \left[1-\operatorname{jinc}^2\left(\frac{D}{2}\frac{z'}{z} \kappa\right)\right] \hfill \\
\quad \times \cos\left[\alpha_k \frac{z'}{z}\left(1-\frac{z'}{z}\right)\kappa^2\right] \text{d}\kappa \text{d}z' + 4\pi^2 \sum\limits_{k=4}^5 c_k \int_0^z\int_0^\infty \kappa \Phi_n\left(\kappa,z'\right) \left[1-\operatorname{jinc}^2\left(\frac{D}{2}\frac{z'}{z} \kappa\right)\right] \hfill \\
\quad \times \cos\left[\alpha_k \frac{z'}{z}\left(1-\frac{z'}{z}\right)\kappa^2\right] J_0\left[\left(1-\frac{z'}{z}\right)\kappa \rho_\text{D}\right] \text{d}\kappa \text{d}z'.\hfill
\end{gathered}    
\end{equation}

Before evaluating \eq{eq:PR9} using Mellin transform techniques, we note that $\Langle \Delta \ell_\text{PR}^2 \Rangle$ consists of two terms.  The first is the sum of the single-wavelength PR OPL variances at $\lambda_{\text{B}}$ and $\lambda_\text{T}$, while the second is the PR OPL covariance at $\lambda_\text{B}$, $\lambda_\text{T}$, and $\rho_\text{D}$.  If $\left|\lambda_\text{B}-\lambda_\text{T}\right|$ or $\rho_\text{D}$ are large, then the covariance term is approximately zero and \eq{eq:PR9} simplifies to the sum of the single-wavelength variances.

\subsection{Evaluating the PR OPD Variance with Mellin Transforms}
We begin by assuming Kolmogorov turbulence and constant $C_n^2$; \eq{eq:PR9} becomes
\begin{equation}\label{eq:PR10}
\begin{gathered}
\Langle \Delta \ell_\text{PR}^2 \Rangle = \Langle \Delta \ell_{\text{PR},\lambda}^2 \Rangle + \Langle \Delta \ell_{\text{PR,COV}}^2 \Rangle, \hfill \\
\Langle \Delta \ell_{\text{PR},\lambda}^2 \Rangle =-2^{14/3} \frac{5}{9} \sqrt{\pi} \Gamma\begin{bmatrix}5/6\\2/3\end{bmatrix} C_n^2 D^{-2} \sum\limits_{k=1}^3 c_k \int_0^z \left(\frac{z'}{z}\right)^{-2} \int_0^\infty \kappa^{-14/3} \hfill \\
\quad \times \left[J_1^2\left(\frac{D}{2}\frac{z'}{z}\kappa\right)-\frac{1}{4}\left(\frac{D}{2}\frac{z'}{z}\kappa\right)^2\right] \cos\left[\alpha_k \frac{z'}{z}\left(1-\frac{z'}{z}\right)\kappa^2\right] \text{d}\kappa \text{d}z' ,\hfill \\
\Langle \Delta \ell_{\text{PR,COV}}^2 \Rangle = -2^{14/3} \frac{5}{9} \sqrt{\pi} \Gamma\begin{bmatrix}5/6\\2/3\end{bmatrix} C_n^2 D^{-2} \sum\limits_{k=4}^5 c_k \int_0^z \left(\frac{z'}{z}\right)^{-2} \int_0^\infty \kappa^{-14/3}\hfill \\
\quad \times  \left[J_1^2\left(\frac{D}{2}\frac{z'}{z}\kappa\right)-\frac{1}{4}\left(\frac{D}{2}\frac{z'}{z}\kappa\right)^2\right]  \cos\left[\alpha_k \frac{z'}{z}\left(1-\frac{z'}{z}\right)\kappa^2\right] J_0\left[\left(1-\frac{z'}{z}\right)\kappa \rho_\text{D}\right] \text{d}\kappa \text{d}z',\hfill
\end{gathered}        
\end{equation}
where the $\Gamma$ notation signifies~\cite{Sasiela:07}
\begin{equation}
\Gamma\begin{bmatrix}a_1,a_2,\cdots,a_n\\b_1,b_2,\cdots,b_m\end{bmatrix} = \frac{\prod\limits_{j=1}^n \Gamma\left(a_j\right)}{\prod\limits_{j=1}^m \Gamma\left(b_j\right)}.    
\end{equation}

\subsubsection{$\Langle \Delta \ell_{\text{PR},\lambda}^2 \Rangle$}
Starting with the single-wavelength variances term $\Langle \Delta \ell_{\text{PR},\lambda}^2 \Rangle$, we make the change of variables $x= \kappa \sqrt{\alpha_k z'\left(1-z'/z\right)/z}$ and simplify yielding
\begin{equation}\label{eq:PR12}
\begin{gathered}
 \Langle \Delta \ell_{\text{PR},\lambda}^2 \Rangle = -2^{14/3} \frac{5}{9} \sqrt{\pi} \Gamma\begin{bmatrix}5/6\\2/3\end{bmatrix} C_n^2 D^{-2} \sum\limits_{k=1}^3 c_k \alpha_k^{11/6} \int_0^z \left(\frac{z'}{z}\right)^{-1/6} \left(1-\frac{z'}{z}\right)^{11/6}  \hfill \\
 \quad \times \int_0^\infty \frac{\text{d}x}{x} x^{-11/3} \cos\left(x^2\right) \left[J_1^2\left(x\frac{D}{2\sqrt{\alpha_k}} \sqrt{\frac{z'/z}{1-z'/z}} \right)-\frac{1}{4}\left( x\frac{D}{2\sqrt{\alpha_k}} \sqrt{\frac{z'/z}{1-z'/z}} \right)^2\right]  \text{d}z'. \hfill
\end{gathered}
\end{equation}
Applying the Mellin convolution theorem~\cite{Sasiela:07,Brychkov2018,Andrews2025} and gamma function identities transforms \eq{eq:PR12} into 
\begin{equation}\label{eq:PR13}
\begin{gathered}
\Langle \Delta \ell_{\text{PR},\lambda}^2 \Rangle = -\frac{4}{\sqrt{\pi}} \frac{5}{9} \Gamma\begin{bmatrix}5/6\\2/3,11/3\end{bmatrix} C_n^2 z D^{-2} \sum\limits_{k=1}^3 c_k \alpha_k^{11/6} \frac{1}{\text{j}2\pi} \int_C \left(\frac{D^2}{8\alpha_k}\right)^{-2s} \Gamma\begin{bmatrix} s+1/2,s+1\end{bmatrix}\hfill \\
\quad \times\, \Gamma\begin{bmatrix}s+23/12\end{bmatrix}   \Gamma\begin{bmatrix} -s-11/12,-s+1/4,-s+3/4,-s+5/12,-s+11/12\\-s+1,-s+3/2,-s+1/2,-s+1\end{bmatrix} \text{d}s, \hfill
\end{gathered}    
\end{equation}
where the integration contour $C$ crosses the real $s$ axis between $-1 < \operatorname{Re}\left(s\right) < -11/12$.  Note that the integral in \eq{eq:PR13} converges for all values of $D^2/\left(8 \alpha_k\right)$ when $C$ is closed to the left.   

The contour integral in \eq{eq:PR13} is in the form of a Meijer G-function~\cite{Wolfram,Sasiela:07, Gradshteyn2000,Brychkov2018,Andrews2025}, namely,
\begin{equation}\label{eq:PR14}
\begin{gathered}
G^{m,n}_{p,q}\left(z\left|\begin{matrix} a_1,\cdots,a_n;a_{n+1},\cdots a_p\\b_1,\cdots,b_m;b_{m+1},\cdots, b_q\end{matrix}\right.\right) \hfill \\
\quad = \frac{1}{\text{j}2\pi} \int_\gamma \frac{\prod_{j=1}^m\Gamma\left(b_j+s\right)\prod_{j=1}^n\Gamma\left(1-a_j-s\right)}{\prod_{j=m+1}^q\Gamma\left(1-b_j-s\right)\prod_{j=n+1}^p\Gamma\left(a_j+s\right)}z^{-s} \text{d}s. \hfill
\end{gathered}
\end{equation}
By definition of the Meijer G-function, the integration contour $\gamma$ in \eq{eq:PR14} passes between the poles arising from the $\Gamma\left(b_j+s\right)$ and $\Gamma\left(1-a_j-s\right)$ gamma functions.  In our particular case, $\gamma$ includes the pole at $s = -1/2$, while $C$ in \eq{eq:PR13} does not.  Therefore, to write \eq{eq:PR13} in terms of a Meijer G-function, we need to subtract the $s=-1/2$ pole contribution from $G_{p,q}^{m,n}$.  Putting \eq{eq:PR13} into the form of \eq{eq:PR14} and applying Cauchy's integral formula~\cite{Arfken2013,Gbur_2011} to find the residue at $s=-1/2$ yields 
\begin{equation}\label{eq:PR15}
\begin{gathered}
\Langle \Delta \ell_{\text{PR},\lambda}^2 \Rangle = -\frac{5}{24} \Gamma\begin{bmatrix} 5/6, -5/6,7/3\\2/3,17/6,23/6\end{bmatrix}C_n^2 z D^{5/3}\hfill \\
\; -\, 2^{-1/3} \frac{\sqrt{\pi}}{3}\Gamma\begin{bmatrix}5/6,17/12,11/6,7/12\\2/3,11/3\end{bmatrix} C_n^2 z \left[\left(\frac{z}{k_\text{B}}\right)^{5/6} + \left(\frac{z}{k_\text{T}}\right)^{5/6} \right]\hfill \\
\; -\, \frac{2}{\sqrt{\pi}} \frac{5}{9} \Gamma\begin{bmatrix} 5/6\\2/3,11/3 \end{bmatrix} C_n^2 z D^{-2} \sum\limits_{k=2}^3 \alpha_k^{11/6} G_{5,7}^{3,5}\left[\left.\left(\frac{D^2}{8\alpha_k}\right)^2\right| \begin{matrix} 23/12,3/4,1/4,7/12,1/12;\text{---}\\1/2,1,23/12;0,-1/2,1/2,0\end{matrix}\right]. \hfill 
\end{gathered}
\end{equation}
We note that the first line of \eq{eq:PR15} is the geometrical optics (GO) PR OPL variance originally derived by \citeA{Noll:76}.  

\subsubsection{$\Langle \Delta \ell_{\text{PR,COV}}^2 \Rangle$}
Proceeding to the covariance term in \eq{eq:PR10} and again making the change of variables $x= \kappa \sqrt{\alpha_k z'\left(1-z'/z\right)/z}$ produces
\begin{equation}\label{eq:PR16}
\begin{gathered}
\Langle \Delta \ell_{\text{PR,COV}}^2 \Rangle = -2^{14/3} \frac{5}{9} \sqrt{\pi} \Gamma\begin{bmatrix}5/6\\2/3\end{bmatrix} C_n^2 D^{-2} \sum\limits_{k=4}^5 c_k \alpha_k^{11/6} \int_0^z \left(\frac{z'}{z}\right)^{-1/6} \left(1-\frac{z'}{z}\right)^{11/6}  \hfill \\
 \quad \times \int_0^\infty \frac{\text{d}x}{x} x^{-11/3} \cos\left(x^2\right)  \left[J_1^2\left(x\frac{D}{2\sqrt{\alpha_k}} \sqrt{\frac{z'/z}{1-z'/z}} \right)-\frac{1}{4}\left( x\frac{D}{2\sqrt{\alpha_k}} \sqrt{\frac{z'/z}{1-z'/z}} \right)^2\right] \hfill \\
 \quad \times J_0\left(x\frac{\rho_\text{D}}{\sqrt{\alpha_k}}\sqrt{\frac{1-z'/z}{z'/z}}\right) \text{d}z'. \hfill  
\end{gathered}    
\end{equation}
Although similar in form to \eq{eq:PR12}, applying the Mellin convolution theorem to \eq{eq:PR16} results in contour integrals in two complex planes due to the additional isoplanatic parameter $\rho_\text{D}/\sqrt{\alpha_k}$.  Following the procedure for evaluating such integrals in \citeA{Sasiela:07} and \citeA{10.1063/1.530086}, we find five groups of so called ``two-poles'' (38 two-poles in total) that, depending on the values of $D^2/\alpha_k$ and $\rho_\text{D}^2/\alpha_k$, contribute to the solution.  The residues of 12 of these 38 two-poles result in infinite series (Taylor series) that converge for all values of $D^2/\alpha_k$ and $\rho_\text{D}^2/\alpha_k$.  The sum of these 12 Taylor series is the solution to the integral; nonetheless, that result is of limited utility since the series are slow to converge for physically relevant values of $D^2/\alpha_k$ and $\rho_\text{D}^2/\alpha_k$.

We, therefore, take a different approach to evaluating \eq{eq:PR16}.  Firstly, note that  
\begin{equation}
\frac{D^2}{\alpha_k} \sim \frac{D^2}{\left(z/k\right)} = \frac{kD^2}{z},
\end{equation}
which is approximately the Fresnel number $N_F$~\cite{IFO,Gaskill1978,Sasiela:07}.  Since our application is beam control or projection, $N_F$ is typically greater than unity and therefore, we are interested in solutions to \eq{eq:PR16} where $D^2/\alpha_k$ is large.  As a result, we do nothing (mathematically) to the bracketed, $J_1^2$ term in \eq{eq:PR16}.  Secondly, and in contrast to $D^2/\alpha_k$, we have no expectation for the value of $\rho_\text{D}$, i.e., both $\rho_\text{D} \to 0$ and $\rho_\text{D} \to \infty$ are physically relevant.  This motivates expanding either $J_0$ (in the case of small $\rho_\text{D}$) or $\operatorname{cosine}$ (for large $\rho_\text{D}$) in Taylor series.  We then apply the Mellin convolution theorem to the remaining integrals.  Because of the Taylor expansions, these integrals are transformed into single contour integrals like \eq{eq:PR13}, which we evaluate using standard complex-plane analysis.

\paragraph{Small $\rho_\text{D}$}
Let us start with small $\rho_\text{D}$.  Expanding $J_0$ in a Taylor series and interchanging the order of the sum and integrals produces
\begin{equation}\label{eq:PR18}
\begin{gathered}
\Langle \Delta \ell_{\text{PR,COV}}^2 \Rangle = -2^{14/3} \frac{5}{9} \sqrt{\pi} \Gamma\begin{bmatrix}5/6\\2/3\end{bmatrix} C_n^2 D^{-2} \sum\limits_{m=0}^\infty \frac{\left(-1\right)^m}{\Gamma^2\left(m+1\right)} \sum\limits_{k=4}^5 c_k \alpha_k^{11/6} \left(\frac{\rho_\text{D}^2}{4\alpha_k}\right)^m \hfill \\
\; \times \int_0^z \left(\frac{z'}{z}\right)^{-1/6-m} \left(1-\frac{z'}{z}\right)^{11/6+m} \int_0^\infty \frac{\text{d}x}{x} x^{-11/3+2m} \cos\left(x^2\right) \hfill \\
\; \times\left[J_1^2\left(x\frac{D}{2\sqrt{\alpha_k}} \sqrt{\frac{z'/z}{1-z'/z}} \right)-\frac{1}{4}\left( x\frac{D}{2\sqrt{\alpha_k}} \sqrt{\frac{z'/z}{1-z'/z}} \right)^2\right] \text{d}z'. \hfill
\end{gathered}    
\end{equation}
Applying the Mellin convolution theorem and using gamma function identities transforms \eq{eq:PR18} into
\begin{equation}\label{eq:PR19}
\begin{gathered}
\Langle \Delta \ell_{\text{PR,COV}}^2 \Rangle = -\frac{4}{\sqrt{\pi}} \frac{5}{9}   \Gamma\begin{bmatrix}5/6\\2/3\end{bmatrix} C_n^2 z D^{-2} \sum\limits_{m=0}^\infty \frac{\left(-1\right)^m}{\Gamma^2\left(m+1\right)} \sum\limits_{k=4}^5 c_k \alpha_k^{11/6} \left(\frac{\rho_\text{D}^2}{2\alpha_k}\right)^m \hfill \\
\; \times \frac{1}{\text{j}2\pi} \int_C \left(\frac{D^2}{8\alpha_k}\right)^{-2s} \Gamma\begin{bmatrix}s+1/2,s+1,s+m/2+17/12,s+m/2+23/12\\s-m/2+17/12\end{bmatrix} \hfill \\
\; \times \Gamma\begin{bmatrix}-s+1/4,-s+3/4,-s+m/2-11/12,-s-m/2+5/12,-s-m/2+11/12\\-s+1,-s+3/2,-s+1/2,-s+1\end{bmatrix} \text{d}s. \hfill
\end{gathered}    
\end{equation}
For $m \leq 2$, the above integral converges for all values $D^2/\left(8 \alpha_k\right)$ when $C$ is closed to the left; it does not converge for $m > 2$.  Like \eq{eq:PR13}, we choose to write \eq{eq:PR19} as a Meijer G-function and again, we must subtract the $s=-1/2$ pole contribution from $G_{p,q}^{m,n}$ to obtain the correct result.  Finding the residue at $s=-1/2$ and using \eq{eq:PR14} yields
\begin{equation}\label{eq:PR20}
\begin{gathered}
\Langle \Delta \ell_{\text{PR,COV}}^2 \Rangle \approx -2^{-13/6} \frac{5}{9} \pi \Gamma\begin{bmatrix}5/6\\2/3,11/3\end{bmatrix} C_n^2 z \sum\limits_{m=0}^{2} \sum\limits_{k=4}^5 \alpha_k^{5/6} \frac{\left[-\rho_\text{D}^2/\left(2\alpha_k\right)\right]^m}{\Gamma^2\left(m+1\right)} \hfill \\
\; \times \Gamma\begin{bmatrix}m+11/6,-m+11/6,m/2-5/12\\-m/2+11/12\end{bmatrix} \hfill \\
\; +\, \frac{4}{\sqrt{\pi}} \frac{5}{9} \Gamma\begin{bmatrix}5/6\\2/3,11/3\end{bmatrix} C_n^2 z D^{-2} \sum\limits_{m=0}^{2} \sum\limits_{k=4}^5 \alpha_k^{11/6} \frac{\left[-\rho_\text{D}^2/\left(2\alpha_k\right)\right]^m}{\Gamma^2\left(m+1\right)} \hfill \\
\; \times G_{6,8}^{4,5}\left[\left(\frac{D^2}{8\alpha_k}\right)^2\left| \begin{matrix} \dfrac{3}{4},\dfrac{1}{4},-\dfrac{m}{2}+\dfrac{23}{12},\dfrac{m}{2}+\dfrac{7}{12},\dfrac{m}{2}+\dfrac{1}{12};-\dfrac{m}{2}+\dfrac{17}{12}\\[10pt]\dfrac{1}{2},1,\dfrac{m}{2}+\dfrac{17}{12},\dfrac{m}{2}+\dfrac{23}{12};0,-\dfrac{1}{2},\dfrac{1}{2},0\end{matrix}\right.\right]. \hfill
\end{gathered}
\end{equation}

\paragraph{Large $\rho_\text{D}$}
Proceeding to large $\rho_\text{D}$, we expand $\cos\left(x^2\right)$ in \eq{eq:PR16} and interchange the sum and integrals yielding
\begin{equation}\label{eq:PR21}
\begin{gathered}
\Langle \Delta \ell_{\text{PR,COV}}^2 \Rangle = -2^{14/3} \frac{5}{9} \sqrt{\pi} \Gamma\begin{bmatrix}5/6\\2/3\end{bmatrix} C_n^2 D^{-2} \sum\limits_{m=0}^{\infty} \frac{\left(-1\right)^m}{\Gamma\left(2m+1\right)} \sum\limits_{k=4}^5 c_k \alpha_k^{11/6} \hfill \\
\; \times \int_0^z \left(\frac{z'}{z}\right)^{-1/6} \left(1-\frac{z'}{z}\right)^{11/6} \int_0^\infty \frac{\text{d}x}{x} x^{-11/3+4m} J_0\left(x \frac{\rho_\text{D}}{\sqrt{\alpha_k}} \sqrt{\frac{1-z'/z}{z'/z}}\right)\hfill \\
\; \times\left[J_1^2\left(x\frac{D}{2\sqrt{\alpha_k}} \sqrt{\frac{z'/z}{1-z'/z}} \right)-\frac{1}{4}\left( x\frac{D}{2\sqrt{\alpha_k}} \sqrt{\frac{z'/z}{1-z'/z}} \right)^2\right] \text{d}z'. \hfill
\end{gathered}    
\end{equation}
Again, applying the Mellin convolution theorem (and gamma function identities) transforms \eq{eq:PR21} into a single contour integral such that
\begin{equation}\label{eq:PR22}
\begin{gathered}
\Langle \Delta \ell_{\text{PR,COV}}^2 \Rangle = -\frac{2^{5/3}}{\pi}\frac{5}{9} \Gamma\begin{bmatrix}5/6\\2/3,11/3\end{bmatrix} C_n^2 z D^{-2} \rho_\text{D}^{11/3} \sum\limits_{m=0}^{\infty} \frac{\left(-1\right)^m}{\Gamma\left(2m+1\right)} \sum\limits_{k=4}^5 c_k \left(\frac{\rho_\text{D}^2}{4 \alpha_k}\right)^{-2m} \hfill \\ \; \times \frac{1}{\text{j}2\pi} \int_C \left(\frac{D}{\rho_\text{D}}\right)^{-2s} \Gamma\begin{bmatrix}s+1,s-m+7/3,s-m+17/6\\s-2m+17/6\end{bmatrix} \hfill \\
\; \times \Gamma\begin{bmatrix}-s+2m-11/6,-s+1/2,-s+m-1/2,-s+m\\-s+2,-s+1\end{bmatrix} \text{d}s. \hfill
\end{gathered}    
\end{equation}
The above integral converges only for $m \leq 1$.  Note that the pole at $s = -1$ lies to the right of $C$ (unlike the other positive $s$ gamma function poles) and therefore, we must subtract the $s = -1$ pole contribution from $G_{p,q}^{m,n}$ to obtain the correct result.  Performing the requisite analysis, we obtain
\begin{equation}\label{eq:PR23}
\begin{gathered}
\Langle \Delta \ell_{\text{PR,COV}}^2 \Rangle \approx -\frac{2^{-1/3}}{\sqrt{\pi}}\frac{5}{9} \Gamma\begin{bmatrix}5/6\\2/3,11/3\end{bmatrix} C_n^2 z \rho_\text{D}^{5/3} \sum\limits_{m=0}^1 \sum\limits_{k=4}^5 \frac{\left(-16\alpha_k^2/\rho_\text{D}^4\right)^m}{\Gamma\left(2m+1\right)} \hfill \\
\quad \times \Gamma\begin{bmatrix}-m+4/3,-m+11/6,2m-5/6,m+1/2,m+1\\-2m+11/6\end{bmatrix} \hfill \\
\quad +\, \frac{2^{5/3}}{\pi}\frac{5}{9} \Gamma\begin{bmatrix}5/6\\2/3,11/3\end{bmatrix} C_n^2 z D^{-2} \rho_\text{D}^{11/3} \sum\limits_{m=0}^1 \sum\limits_{k=4}^5 \frac{\left(-16\alpha_k^2/\rho_\text{D}^4\right)^m}{\Gamma\left(2m+1\right)} \hfill \\
\quad \times G_{5,5}^{3,4}\left[\left(\frac{D}{\rho_\text{D}}\right)^2\left| \begin{matrix} \dfrac{1}{2},-2m+\dfrac{17}{6},-m+\dfrac{3}{2},-m+1;-2m+\dfrac{17}{6}\\[10pt]1,-m+\dfrac{7}{3},-m+\dfrac{17}{6};0,-1\end{matrix}\right.\right]. \hfill
\end{gathered}    
\end{equation}
\begin{figure}[t]
    \centering
    \includegraphics[width=\linewidth]{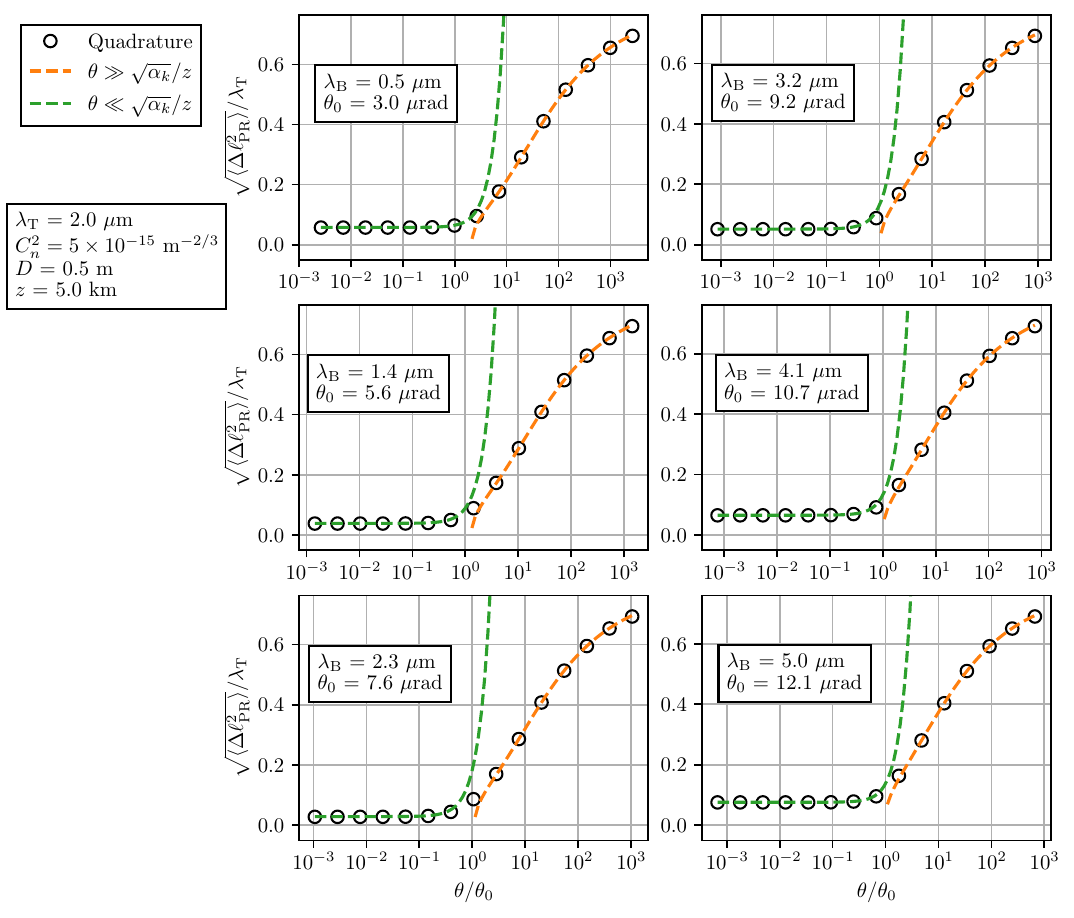}
    \caption{Normalized two-wavelength PR OPD error versus $\theta/\theta_0$ at $\lambda_\text{T} = 2 \text{ } \mu\text{m}$ and several values of $\lambda_\text{B}$.}
    \label{fig:PR}
\end{figure}

\subsection{Summary and Discussion}Combining the above results, the anisoplanatic two-wavelength PR OPD variance is
\begin{equation}\label{eq:PR24}
\begin{gathered}
 \Langle \Delta \ell_\text{PR}^2 \Rangle = \Langle \Delta \ell_{\text{PR},\lambda}^2 \Rangle + \Langle \Delta \ell_{\text{PR,COV}}^2 \Rangle, \hfill \\
 \Langle \Delta \ell_{\text{PR},\lambda}^2 \Rangle = \text{\eq{eq:PR15}}, \hfill \\
 \Langle \Delta \ell_{\text{PR,COV}}^2 \Rangle \approx \begin{cases} \text{\eq{eq:PR20}} & {D^2}/\alpha_k \gg 1,\, \rho_\text{D}^2/\alpha_k \ll 1 \\
 \text{\eq{eq:PR23}} & {D^2}/\alpha_k \gg 1,\, \rho_\text{D}^2/\alpha_k \gg 1 \end{cases}. \hfill
\end{gathered}    
\end{equation}

Figure~\ref{fig:PR} shows plots of the two-wavelength PR OPD error normalized by $\lambda_\text{T} = 2 \text{ } \mu\text{m}$ versus $\theta/\theta_0$ for several values of $\lambda_\text{B}$, where $\theta = \rho_\text{D}/z$ and $\theta_0$ is the two-wavelength isoplanatic angle derived in \citeA{si1}:
\begin{equation}\label{eq:2wvltheta0}
\theta_0 = \left(-\frac{\sqrt{\pi}}{8} \Gamma\begin{bmatrix}-5/6\\2/3\end{bmatrix} C_n^2 k_\text{B} k_\text{T} z^{8/3}\right)^{-3/5}.    
\end{equation}
The values for $C_n^2$, $D$, $z$, and $\theta_0$ are annotated on the figure.  The ``Quadrature'' results are from evaluating \eq{eq:PR10} numerically using adaptive quadrature.  

We observe excellent agreement between the Quadrature and asymptotic results over the latter's applicable regions.  Although it does not explicitly appear in \eq{eq:PR24}, the two-wavelength isoplanatic angle $\theta_0$ is the inflection point of the curves.  This makes physical sense: For $\theta < \theta_0$, the fields emitted by the two point sources in Figure~\ref{fig:geom} are angularly correlated, and the difference between $\lambda_\text{B}$ and $\lambda_\text{T}$ is the primary source of the OPD error.  Consequently, we observe near constant $\sqrt{\Langle \Delta \ell_\text{PR}^2\Rangle}/\lambda_\text{T}$ in these regions of the curves.  In addition, $\sqrt{\Langle \Delta \ell_\text{PR}^2\Rangle}/\lambda_\text{T}$ initially decreases as $\lambda_\text{B}$ approaches $\lambda_\text{T} = 2 \text{ } \mu\text{m}$ (first column of plots), then increases as $\left|\lambda_\text{B}-\lambda_\text{T}\right|$ grows (second column of plots). 

In contrast, when $\theta > \theta_0$, the fields emitted by the two point sources are angularly decorrelated.  Under this condition, the OPD error is a combination of the beacon and transmitter wavelength and angular separations.  As mentioned earlier, at large wavelength or angular separations, the OPD variance simplifies to the sum of the single-wavelength variances.  This explains why the PR OPD errors in Figure~\ref{fig:PR} approach roughly the same value for $\theta/\theta_0 \gg 1$.     

\section{Anisoplanatic, Two-Wavelength PTR OPD Variance}
To derive the two-wavelength PTR OPD variance with anisoplanatism $\Langle \Delta \ell_{\text{PTR}}^2\Rangle$, we first need an expression for the tilt OPD variance.  Because of the way the tilt OPD variance is formulated, there is no added complexity in obtaining the Zernike-mode OPD variance---i.e., the OPD variance for any optical aberration greater than piston---and specializing that result to tilt.  This is the approach we take here.  Much of the mathematics is similar to that presented above, and therefore, we omit many of the details for the sake of brevity.

Our analysis begins with the OPL expanded in terms of Zernike polynomials, namely,
\begin{equation}\label{eq:PTR1}
\ell\left(\boldsymbol{\rho},z;\lambda,\boldsymbol{\rho}_c\right) = \sum\limits_{i,j} a_{i,j}\left(\lambda,\boldsymbol{\rho}_c\right) Z_{i,j}\left(\frac{\rho}{D/2},\phi\right), 
\end{equation}
where $i,j$ are the radial and azimuthal indices of the Zernike polynomial $Z_{i,j}$~\cite{Noll:76,OIAWA,Lakshminarayanan10042011,Sasiela:07} and $a_{i,j}$ is the weight/coefficient of $Z_{i,j}$ in meters.  Because Zernike polynomials are orthonormal over circular apertures,
\begin{equation}\label{eq:PTR2}
\ell_{i,j}\left(\boldsymbol{\rho},z;\lambda,\boldsymbol{\rho}_c\right) = a_{i,j}\left(\lambda,\boldsymbol{\rho}_c\right) Z_{i,j}\left(\frac{\rho}{D/2},\phi\right)
\end{equation}
and
\begin{equation}\label{eq:PTR3}
a_{i,j}\left(\lambda,\boldsymbol{\rho}_c\right) = \frac{1}{A}\iint_{-\infty}^{\infty} \Circ{\frac{\boldsymbol{\rho}}{D/2}} Z_{i,j}\left(\frac{\rho}{D/2},\phi\right) \ell\left(\boldsymbol{\rho},z;\lambda,\boldsymbol{\rho}_c\right) \text{d}^2\rho.    
\end{equation}
Using \eqs{eq:PTR2}{eq:PTR3}, it follows that the anisotropic, two-wavelength Zernike-mode OPD variance takes the form
\begin{equation}\label{eq:PTR4}
\begin{split}
\Langle \Delta \ell_{Z_{i,j}}^2 \Rangle &= \frac{1}{A} \iint_{-\infty}^{\infty} \Circ{\frac{\boldsymbol{\rho}}{D/2}} \Langle \left[\ell_{i,j}\left(\boldsymbol{\rho},z;\lambda_\text{B},\boldsymbol{\rho}_\text{B}\right) - \ell_{i,j}\left(\boldsymbol{\rho},z;\lambda_\text{T},\boldsymbol{\rho}_\text{T}\right)\right]^2\Rangle \text{d}^2\rho \\
&= \Langle a_{i,j}^2\left(\lambda_\text{B},\boldsymbol{\rho}_\text{B}\right)\Rangle + \Langle a_{i,j}^2\left(\lambda_\text{T},\boldsymbol{\rho}_\text{T}\right)\Rangle - 2 \Langle a_{i,j}\left(\lambda_\text{B},\boldsymbol{\rho}_\text{B}\right)a_{i,j}\left(\lambda_\text{T},\boldsymbol{\rho}_\text{T}\right)\Rangle.
\end{split}
\end{equation}
We now focus on the Zernike coefficient covariance [last moment in \eq{eq:PTR4}], as the single-wavelength, single-location variances are simply special cases of it.

Returning briefly to \eq{eq:PTR3} and introducing the Fourier transform of $Z_{i,j}$, we obtain
\begin{equation}\label{eq:PTR5}
a_{i,j}\left(\lambda,\boldsymbol{\rho}_c\right) = \frac{\pi}{A}\iint_{-\infty}^{\infty} Q_{i,j}\left(\boldsymbol{f}\right) \iint_{-\infty}^{\infty} \ell\left(\boldsymbol{\rho},z;\lambda,\boldsymbol{\rho}_c\right) \exp\left(\text{j}2\pi\boldsymbol{f}\cdot \frac{\boldsymbol{\rho}}{D/2}\right) \text{d}^2\rho \text{d}^2 f,    
\end{equation}
where $Q_{i,j}$ is the Fourier transform of $Z_{i,j}$ given in \citeA{Noll:76} and \citeA{Sasiela:07}.  Substituting \eq{eq:PTR5} into the Zernike coefficient covariance in \eq{eq:PTR4} and simplifying yields
\begin{equation}\label{eq:PTR6}
\begin{gathered}
\Langle a_{i,j}\left(\lambda_\text{B},\boldsymbol{\rho}_\text{B}\right)a_{i,j}\left(\lambda_\text{T},\boldsymbol{\rho}_\text{T}\right)\Rangle = \frac{\pi}{A} \iint_{-\infty}^{\infty}\left|Q_{i,j}\left(\boldsymbol{f}\right)\right|^2 \hfill \\
\quad \times \iint_{-\infty}^{\infty}\frac{1}{k_\text{B} k_\text{T}}B_S\left(\boldsymbol{\rho},z;\lambda_\text{B},\boldsymbol{\rho}_\text{B};\lambda_\text{T},\boldsymbol{\rho}_\text{T}\right) \exp\left(\text{j}2\pi\boldsymbol{f}\cdot \frac{\boldsymbol{\rho}}{D/2}\right) \text{d}^2\rho \text{d}^2 f.    
\end{gathered}
\end{equation}
Recall that $B_S$ is the phase covariance function derived in \citeA{si1}.  

Substituting $B_S$ from \citeA{si1} [see Eq.~(14) from the Supporting Information] as well as $Q_{i,j}$ from \citeA{Noll:76} [or \citeA{Sasiela:07}] in \eq{eq:PTR6}, rearranging the integrals, and lastly, evaluating the trivial Dirac delta function integrals reveals
\begin{equation}\label{eq:PTR7}
\begin{gathered}
 \Langle a_{i,j}\left(\lambda_\text{B},\boldsymbol{\rho}_\text{B}\right)a_{i,j}\left(\lambda_\text{T},\boldsymbol{\rho}_\text{T}\right)\Rangle = \frac{\pi}{2} \left(i+1\right) \sum\limits_{k=4}^5 \int_0^z \int_0^\infty \kappa \Phi_n\left(\kappa,z'\right)  \hfill \\
 \quad \times \left\{\frac{2 J_{i+1}\left[D z' \kappa/\left(2z\right)\right]}{ D z' \kappa/\left(2z\right) }\right\}^2   \exp\left[\text{j}\alpha_k\frac{z'}{z}\left(1-\frac{z'}{z}\right)\kappa^2\right] \hfill \\
 \quad \times \int_0^{2\pi} \begin{Bmatrix}2 \cos^2\left(2j \varphi\right)\\ 2 \sin^2\left(2j \varphi\right) \\ 1 \end{Bmatrix} \exp\left[-\text{j}\left(1-\frac{z'}{z}\right)\kappa \rho_\text{D} \cos\left(\varphi-\phi_\text{D}\right)\right] \text{d} \varphi \text{d}\kappa \text{d}z' + \text{c.c.},  \hfill
\end{gathered}    
\end{equation}
where $\phi_\text{D}$ is the angle of $\boldsymbol{\rho}_\text{D}$.  The integral over $\varphi$ links the orientation of the point sources [see Figure~\ref{fig:geom}] to Zernike-mode symmetry.  In the braces, the top entry/element corresponds to an $x$ mode, i.e., $x$ tilt, $x$ astigmatism, $x$ coma, etcetera; the middle element to a $y$ mode; and finally, the third element to a rotationally invariant $j=0$ mode.  This explains why the $\varphi$ integral does not appear in $\Langle \ell_\text{PR}^2 \Rangle$, as piston $Z_{0,0}$ is $\phi$ invariant.  

Using variable substitution and Bessel function integral identities, the integral over $\varphi$ equals
\begin{equation}\label{eq:PTR8}
\begin{gathered}
\int_0^{2\pi} \begin{Bmatrix}2 \cos^2\left(2j \varphi\right)\\ 2 \sin^2\left(2j \varphi\right) \\ 1 \end{Bmatrix} \exp\left[-\text{j}\left(1-\frac{z'}{z}\right)\kappa \rho_\text{D} \cos\left(\varphi-\phi_\text{D}\right)\right] \text{d} \varphi \hfill \\
\quad = 2\pi J_0\left[\left(1-\frac{z'}{z}\right) \kappa \rho_\text{D}\right] + 2\pi \begin{Bmatrix}1\\ -1\\ 0 \end{Bmatrix} \left(-1\right)^j J_{2j}\left[\left(1-\frac{z'}{z}\right) \kappa \rho_\text{D}\right] \cos\left(2j\phi_\text{D}\right). \hfill
\end{gathered}
\end{equation}
Inserting \eq{eq:PTR8} into \eq{eq:PTR7} yields the Zernike coefficient covariance:
\begin{equation}\label{eq:PTR9}
\begin{gathered}
 \Langle a_{i,j}\left(\lambda_\text{B},\boldsymbol{\rho}_\text{B}\right)a_{i,j}\left(\lambda_\text{T},\boldsymbol{\rho}_\text{T}\right)\Rangle = 2\pi^2 \left(i+1\right) \sum\limits_{k=4}^5 \int_0^z \int_0^\infty \kappa \Phi_n\left(\kappa,z'\right)  \hfill \\
 \quad \times \left\{\frac{2 J_{i+1}\left[D z' \kappa/\left(2z\right)\right]}{ D z' \kappa/\left(2z\right) }\right\}^2   \cos\left[\alpha_k\frac{z'}{z}\left(1-\frac{z'}{z}\right)\kappa^2\right] \left\{J_0\left[\left(1-\frac{z'}{z}\right) \kappa \rho_\text{D}\right] \right. \hfill \\
 \quad +\left. \begin{Bmatrix}1\\ -1\\ 0 \end{Bmatrix} \left(-1\right)^j J_{2j}\left[\left(1-\frac{z'}{z}\right) \kappa \rho_\text{D}\right] \cos\left(2j\phi_\text{D}\right) \right\} \text{d}\kappa \text{d}z'. \hfill
\end{gathered}    
\end{equation}

Lastly, we obtain the anisoplanatic, two-wavelength Zernike-mode OPD variance by specializing \eq{eq:PTR9} to find $\Langle a_{i,j}^2\left(\lambda,\boldsymbol{\rho}_c\right)\Rangle$ and then substituting everything into \eq{eq:PTR4}.  The result is
\begin{equation}\label{eq:PTR10}
\begin{gathered}
\Langle \Delta \ell_{Z_{i,j}}^2 \Rangle = \Langle \Delta \ell_{Z_{i,j},\lambda}^2 \Rangle + \Langle \Delta \ell_{Z_{i,j},\text{COV}}^2\Rangle   , \hfill \\ 
\Langle \Delta \ell_{Z_{i,j},\lambda}^2 \Rangle = 4\pi^2 \sum\limits_{k=1}^3 c_k \int_0^z\int_0^\infty \kappa \Phi_n\left(\kappa,z'\right) \left\{\frac{2 J_{i+1}\left[D z' \kappa/\left(2z\right)\right]}{ D z' \kappa/\left(2z\right) }\right\}^2  \hfill \\
\quad \times \cos\left[\alpha_k \frac{z'}{z}\left(1-\frac{z'}{z}\right)\kappa^2\right] \text{d}\kappa \text{d}z' \hfill \\
\Langle \Delta \ell_{Z_{i,j},\text{COV}}^2\Rangle = 4\pi^2 \sum\limits_{k=4}^5 c_k \int_0^z\int_0^\infty \kappa \Phi_n\left(\kappa,z'\right) \left\{\frac{2 J_{i+1}\left[D z' \kappa/\left(2z\right)\right]}{ D z' \kappa/\left(2z\right) }\right\}^2  \hfill \\
\quad \times \cos\left[\alpha_k \frac{z'}{z}\left(1-\frac{z'}{z}\right)\kappa^2\right] \left\{J_0\left[\left(1-\frac{z'}{z}\right) \kappa \rho_\text{D}\right] \right. \hfill \\
 \quad +\left. \begin{Bmatrix}1\\ -1\\ 0 \end{Bmatrix} \left(-1\right)^j J_{2j}\left[\left(1-\frac{z'}{z}\right) \kappa \rho_\text{D}\right] \cos\left(2j\phi_\text{D}\right) \right\} \text{d}\kappa \text{d}z'. \hfill
\end{gathered}    
\end{equation}
We now proceed to evaluating $\Langle \Delta \ell_{Z_{i,j},\lambda}^2 \Rangle$ and $\Langle \Delta \ell_{Z_{i,j},\text{COV}}^2\Rangle $ using Mellin transform techniques.

\subsection{$\Langle \Delta \ell_{Z_{i,j},\lambda}^2 \Rangle$}
Assuming Kolmogorov turbulence and constant $C_n^2$ as well as making the change of variables $x= \kappa \sqrt{\alpha_k z'\left(1-z'/z\right)/z}$ yields
\begin{equation}\label{eq:PTR11}
\begin{gathered}
\Langle \Delta \ell_{Z_{i,j},\lambda}^2 \Rangle = 2^{14/3} \frac{5}{9} \sqrt{\pi} \Gamma\begin{bmatrix}5/6\\2/3\end{bmatrix} \left(i+1\right) C_n^2 D^{-2} \sum\limits_{k=1}^3 c_k \alpha_k^{11/6} \int_0^z \left(\frac{z'}{z}\right)^{-1/6} \left(1-\frac{z'}{z}\right)^{11/6}  \hfill \\
 \quad \times \int_0^\infty \frac{\text{d}x}{x} x^{-11/3} \cos\left(x^2\right) J_{i+1}^2\left(x\frac{D}{2\sqrt{\alpha_k}} \sqrt{\frac{z'/z}{1-z'/z}} \right) \text{d}z'. \hfill    
\end{gathered}    
\end{equation}
The Mellin convolution theorem transforms \eq{eq:PTR11} into the following contour integral:
\begin{equation}\label{eq:PTR12}
\begin{gathered}
\Langle \Delta \ell_{Z_{i,j},\lambda}^2 \Rangle = \frac{4}{\sqrt{\pi}} \frac{5}{9} \Gamma\begin{bmatrix}5/6\\2/3,11/3\end{bmatrix} \left(i+1\right) C_n^2 z D^{-2} \sum\limits_{k=1}^3 c_k \alpha_k^{11/6} \hfill \\
\quad \times \frac{1}{\text{j}2\pi} \int_C \left(\frac{D^2}{8 \alpha_k}\right)^{-2s} \Gamma\begin{bmatrix}s+i/2+1/2,s+i/2+1,s+23/12\end{bmatrix} \hfill \\
\quad \times \Gamma\begin{bmatrix}-s-11/12,-s+1/4,-s+5/12,-s+3/4,-s+11/12\\-s+i/2+1,-s+i/2+3/2,-s+1/2,-s+1\end{bmatrix} \text{d}s, \hfill
\end{gathered}    
\end{equation}
where $C$ crosses the real $s$ axis between $-\left(i+1\right)/2 < \operatorname{Re}\left(s\right) < -11/12$ with $i \geq 1$.  Unlike the corresponding $\Langle \Delta \ell_{\text{PR},\lambda}^2 \Rangle$ integral in \eq{eq:PR13}, here, $C$ and the Meijer G-function contour $\gamma$ [see \eq{eq:PR14}] are consistent.  Consequently,
\begin{equation}\label{eq:PTR13}
\begin{gathered}
\Langle \Delta \ell_{Z_{i,j},\lambda}^2 \Rangle = \frac{5}{24} \Gamma\begin{bmatrix}5/6,i-5/6,7/3\\2/3,17/6,i+23/6\end{bmatrix}\left(i+1\right)C_n^2 z D^{5/3} \hfill \\
\quad +\, \frac{2}{\sqrt{\pi}} \frac{5}{9} \Gamma\begin{bmatrix}5/6\\2/3,11/3\end{bmatrix} \left(i+1\right) C_n^2 z D^{-2} \sum\limits_{k=2}^3 \alpha_k^{11/6} \hfill \\
\quad \times\, G_{5,7}^{3,5}\left[\left(\frac{D^2}{8\alpha_k}\right)^2\left| \begin{matrix} \dfrac{23}{12},\dfrac{3}{4},\dfrac{1}{4},\dfrac{7}{12},\dfrac{1}{12};\text{---}\\[10pt]\dfrac{i}{2}+\dfrac{1}{2},\dfrac{i}{2}+1,\dfrac{23}{12};-\dfrac{i}{2},-\dfrac{i}{2}-\dfrac{1}{2},\dfrac{1}{2},0\end{matrix}\right.\right]. \hfill 
\end{gathered}   
\end{equation}
The first line of \eq{eq:PTR13} is the GO Zernike-mode OPL variance first derived by \citeA{Noll:76}.

\subsection{$\Langle \Delta \ell_{Z_{i,j},\text{COV}}^2 \Rangle$}
Progressing to the covariance term in \eq{eq:PTR10}, we again assume Kolomogorov turbulence, constant $C_n^2$, and make the variable substitution $x= \kappa \sqrt{\alpha_k z'\left(1-z'/z\right)/z}$ yielding
\begin{equation}\label{eq:PTR14}
\begin{gathered}
\Langle \Delta \ell_{Z_{i,j},\text{COV}}^2 \Rangle =  2^{14/3} \frac{5}{9} \sqrt{\pi} \Gamma\begin{bmatrix}5/6\\2/3\end{bmatrix} \left(i+1\right) C_n^2 D^{-2} \sum\limits_{k=4}^5 c_k \alpha_k^{11/6} \int_0^z \left(\frac{z'}{z}\right)^{-1/6} \left(1-\frac{z'}{z}\right)^{11/6}  \hfill \\
 \quad \times \int_0^\infty \frac{\text{d}x}{x} x^{-11/3} \cos\left(x^2\right) J_{i+1}^2\left(x\frac{D}{2\sqrt{\alpha_k}} \sqrt{\frac{z'/z}{1-z'/z}} \right) \left[J_0\left(x\frac{\rho_\text{D}}{\sqrt{\alpha_k}}\sqrt{\frac{1-z'/z}{z'/z}}\right) \right. \hfill \\
 \quad +\left. \begin{Bmatrix}1\\-1\\0\end{Bmatrix}\left(-1\right)^j J_{2j}\left(x\frac{\rho_\text{D}}{\sqrt{\alpha_k}}\sqrt{\frac{1-z'/z}{z'/z}}\right) \cos\left(2j\phi_\text{D}\right) \right] \text{d}z'. \hfill       
\end{gathered}    
\end{equation}
Hereafter, we focus on evaluating
\begin{equation}\label{eq:PTR15}
\begin{gathered}
I_{i,j} = \int_0^z \left(\frac{z'}{z}\right)^{-1/6} \left(1-\frac{z'}{z}\right)^{11/6}  \int_0^\infty \frac{\text{d}x}{x} x^{-11/3} \cos\left(x^2\right) \hfill \\
 \quad \times J_{2j}\left(x\frac{\rho_\text{D}}{\sqrt{\alpha_k}}\sqrt{\frac{1-z'/z}{z'/z}}\right) J_{i+1}^2\left(x\frac{D}{2\sqrt{\alpha_k}} \sqrt{\frac{z'/z}{1-z'/z}} \right) \text{d}z'. \hfill    
\end{gathered}
\end{equation}
We specialize or substitute $I_{i,j}$ into \eq{eq:PTR14} to obtain $\Langle \Delta \ell_{Z_{i,j},\text{COV}}^2 \Rangle$ when relevant.  Like $\Langle \Delta \ell_{\text{PR},\text{COV}}^2 \Rangle$ in \eq{eq:PR16}, we seek solutions to $I_{i,j}$ where ${D^2}/\alpha_k \gg 1$ and for small and large $\rho_\text{D}$.  We start with the former.

\subsubsection{Small $\rho_\text{D}$}
Expanding $J_{2j}$ in a Taylor series, applying the Mellin convolution theorem, and writing the contour integral as a Meijer G-function produces
\begin{equation}\label{eq:PTR16}
\begin{gathered}
I_{i,j} \approx \frac{2^{-8/3}}{\Gamma\left(11/3\right)} \frac{z}{\pi} \left(\frac{\rho_\text{D}^2}{2\alpha_k}\right)^j \sum\limits_{m=0}^M \frac{\left[-\rho_\text{D}^2/\left(2\alpha_k\right)\right]^m}{\Gamma\begin{bmatrix}m+1,m+2j+1\end{bmatrix}} \hfill \\
\quad \times G_{6,8}^{4,5} \left[\left(\frac{D^2}{8\alpha_k}\right)^2\left| \begin{matrix}\dfrac{3}{4},\dfrac{1}{4},-\dfrac{m}{2}-\dfrac{j}{2}+\dfrac{23}{12},\dfrac{m}{2}+\dfrac{j}{2}+\dfrac{7}{12},\dfrac{m}{2}+\dfrac{j}{2}+\dfrac{1}{12};-\dfrac{m}{2}-\dfrac{j}{2}+\dfrac{17}{12}\\[10pt]\dfrac{i}{2}+\dfrac{1}{2},\dfrac{i}{2}+1,\dfrac{m}{2}+\dfrac{j}{2}+\dfrac{17}{12},\dfrac{m}{2}+\dfrac{j}{2}+\dfrac{23}{12};-\dfrac{i}{2},-\dfrac{i}{2}-\dfrac{1}{2},\dfrac{1}{2},0\end{matrix}\right.\right], \hfill
\end{gathered}    
\end{equation}
where the number of terms $M$ for which the contour integral converges depends on $i,j$.  Our ultimate interest is tilt, i.e., $i,j = 1$, and $M = 1$.  After substituting \eq{eq:PTR16} into \eq{eq:PTR14} and simplifying, we obtain 
\begin{equation}\label{eq:PTR17}
    \begin{gathered}
\Langle \Delta \ell_{Z_{i,j},\text{COV}}^2 \Rangle = -\frac{4}{\sqrt{\pi}} \frac{5}{9} \Gamma\begin{bmatrix}5/6\\2/3,11/3\end{bmatrix} \left(i+1\right) C_n^2 z D^{-2} \sum\limits_{m=0}^M \sum\limits_{k=4}^5 \alpha_k^{11/6} \left\{ \frac{\left[-\rho_\text{D}^2/\left(2\alpha_k\right)\right]^m}{\Gamma^2\left(m+1\right)} \right. \hfill \\
\left.\; \times G_{6,8}^{4,5} \left[\left(\frac{D^2}{8\alpha_k}\right)^2\left| \begin{matrix}\dfrac{3}{4},\dfrac{1}{4},-\dfrac{m}{2}+\dfrac{23}{12},\dfrac{m}{2}+\dfrac{7}{12},\dfrac{m}{2}+\dfrac{1}{12};-\dfrac{m}{2}+\dfrac{17}{12}\\[10pt]\dfrac{i}{2}+\dfrac{1}{2},\dfrac{i}{2}+1,\dfrac{m}{2}+\dfrac{17}{12},\dfrac{m}{2}+\dfrac{23}{12};-\dfrac{i}{2},-\dfrac{i}{2}-\dfrac{1}{2},\dfrac{1}{2},0\end{matrix}\right.\right] \right. \hfill \\
\left.\; + \begin{Bmatrix}1\\-1\\0\end{Bmatrix} \frac{\left[-\rho_\text{D}^2/\left(2\alpha_k\right)\right]^{m+j}}{\Gamma\begin{bmatrix}m+1,m+2j+1\end{bmatrix}} \cos\left(2j\phi_\text{D}\right) \right. \hfill \\
\left.\; \times G_{6,8}^{4,5} \left[\left(\frac{D^2}{8\alpha_k}\right)^2\left| \begin{matrix}\dfrac{3}{4},\dfrac{1}{4},-\dfrac{m}{2}-\dfrac{j}{2}+\dfrac{23}{12},\dfrac{m}{2}+\dfrac{j}{2}+\dfrac{7}{12},\dfrac{m}{2}+\dfrac{j}{2}+\dfrac{1}{12};-\dfrac{m}{2}-\dfrac{j}{2}+\dfrac{17}{12}\\[10pt]\dfrac{i}{2}+\dfrac{1}{2},\dfrac{i}{2}+1,\dfrac{m}{2}+\dfrac{j}{2}+\dfrac{17}{12},\dfrac{m}{2}+\dfrac{j}{2}+\dfrac{23}{12};-\dfrac{i}{2},-\dfrac{i}{2}-\dfrac{1}{2},\dfrac{1}{2},0\end{matrix}\right.\right]
\right\}.\hfill
\end{gathered}
\end{equation}

\subsubsection{Large $\rho_\text{D}$}
Returning to \eq{eq:PTR15}, expanding $\cos\left(x^2\right)$ in a Taylor series, and again, applying the Mellin convolution theorem and \eq{eq:PR14} reveals
\begin{equation}\label{eq:PTR18}
\begin{gathered}
I_{i,j} \approx  \frac{2^{8/3}}{\Gamma\left(11/3\right)} \frac{z}{\pi^{3/2}} \left(\frac{\rho_\text{D}^2}{4 \alpha_k}\right)^{11/6}\sum\limits_{m=0}^M \frac{\left(-16 \alpha_k^2/\rho_\text{D}^4\right)^m}{\Gamma\left(2m+1\right)} \hfill \\
\quad \times G_{5,5}^{3,4} \left[\left(\frac{D}{\rho_\text{D}}\right)^2\left| \begin{matrix}\dfrac{1}{2},-2m-j+\dfrac{17}{6},-m+\dfrac{3}{2},-m+1;-2m+j+\dfrac{17}{6}\\[10pt]i+1,-m+\dfrac{7}{3},-m+\dfrac{17}{6};0,-i-1\end{matrix}\right.\right], \hfill
\end{gathered}
\end{equation}
where again $M$ depends on $i,j$.  In the case of tilt, $M = 2$.  Inserting \eq{eq:PTR18} into \eq{eq:PTR14} and simplifying produces
\begin{equation}\label{eq:PTR19}
\begin{gathered}
\Langle \Delta \ell_{Z_{i,j},\text{COV}}^2 \Rangle = -\frac{2^{5/3}}{\pi}\frac{5}{9} \Gamma\begin{bmatrix}5/6\\2/3,11/3\end{bmatrix} \left(i+1\right) C_n^2 z D^{-2} \rho_\text{D}^{11/3} \sum\limits_{m=0}^M \sum\limits_{k=4}^5 \frac{\left(-16 \alpha_k^2/\rho_\text{D}^4\right)^m}{\Gamma\left(2m+1\right)} \hfill \\ 
\; \left\{ G_{5,5}^{3,4} \left[\left(\frac{D}{\rho_\text{D}}\right)^2\left| \begin{matrix}\dfrac{1}{2},-2m+\dfrac{17}{6},-m+\dfrac{3}{2},-m+1;-2m+\dfrac{17}{6}\\[10pt]i+1,-m+\dfrac{7}{3},-m+\dfrac{17}{6};0,-i-1\end{matrix}\right.\right] + \begin{Bmatrix}1\\-1\\0\end{Bmatrix} \left(-1\right)^j \cos\left(2j\phi_\text{D}\right) \right. \hfill \\
\; \left.\times G_{5,5}^{3,4} \left[\left(\frac{D}{\rho_\text{D}}\right)^2\left| \begin{matrix}\dfrac{1}{2},-2m-j+\dfrac{17}{6},-m+\dfrac{3}{2},-m+1;-2m+j+\dfrac{17}{6}\\[10pt]i+1,-m+\dfrac{7}{3},-m+\dfrac{17}{6};0,-i-1\end{matrix}\right.\right] \right\}. \hfill
\end{gathered}    
\end{equation}

\subsection{Summary and Discussion}
In summary, the anisoplanatic, two-wavelength Zernike-mode OPD variance is
\begin{equation}\label{eq:PTR20}
 \begin{gathered}
 \Langle \Delta \ell_{Z_{i,j}}^2 \Rangle = \Langle \Delta \ell_{Z_{i,j},\lambda}^2 \Rangle + \Langle \Delta \ell_{Z_{i,j},\text{COV}}^2 \Rangle, \hfill \\
 \Langle \Delta \ell_{Z_{i,j},\lambda}^2 \Rangle = \text{\eq{eq:PTR13}}, \hfill \\
 \Langle \Delta \ell_{Z_{i,j},\text{COV}}^2 \Rangle \approx \begin{cases} \text{\eq{eq:PTR17}} & {D^2}/\alpha_k \gg 1,\, \rho_\text{D}^2/\alpha_k \ll 1 \\
 \text{\eq{eq:PTR19}} & {D^2}/\alpha_k \gg 1,\, \rho_\text{D}^2/\alpha_k \gg 1 \end{cases}. \hfill
\end{gathered}       
\end{equation}
Furthermore, the anisoplanatic, two-wavelength PTR OPD variance is
\begin{equation}\label{eq:PTR21}
\Langle \Delta \ell_\text{PTR}^2 \Rangle = \Langle \Delta \ell_\text{PR}^2 \Rangle - \left[\Langle \Delta \ell_{Z_{1,1}^x}^2 \Rangle + \Langle \Delta \ell_{Z_{1,1}^y}^2 \Rangle \right],      
\end{equation}
where the superscript $x$ on $Z_{1,1}^x$ signifies $x$ tilt and similarly for $Z_{1,1}^y$.  Note that adding the $x$ and $y$ tilt variances in \eq{eq:PTR21} cancels the $\cos\left(2j\phi_\text{D}\right)$ terms in \eqs{eq:PTR17}{eq:PTR19}, thereby, making $\Langle \Delta \ell_\text{PTR}^2 \Rangle$ independent of point-source orientation.
\begin{figure}[t]
    \centering
    \includegraphics[width=\linewidth]{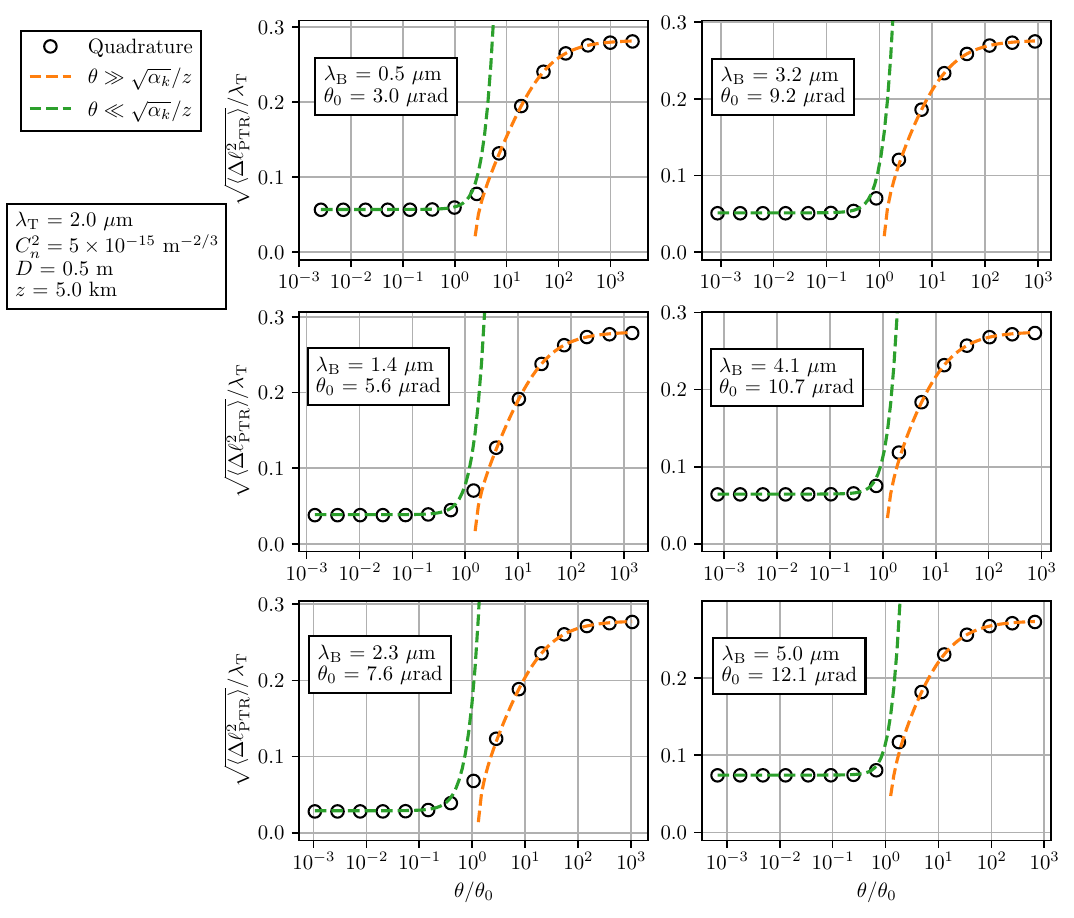}
    \caption{Normalized PTR OPD error versus $\theta/\theta_0$ at $\lambda_\text{T} = 2 \text{ } \mu\text{m}$ and several values of $\lambda_\text{B}$.}
    \label{fig:PTR}
\end{figure}

Figure~\ref{fig:PTR} plots the normalized, two-wavelength PTR OPD error, i.e., $\sqrt{\Langle \Delta \ell_\text{PTR}^2 \Rangle}/\lambda_\text{T}$, versus $\theta/\theta_0$ with $\lambda_\text{T} = 2 \text{ } \mu\text{m}$ and at several values of $\lambda_\text{B}$.  Like in Figure~\ref{fig:PR}, we observe excellent agreement between the Quadrature and asymptotic results (when applicable).  Also, the two-wavelength isoplanatic angle continues to be a strong predictor of the wavefront error.  In other words, when $\theta < \theta_0$, the difference between $\lambda_\text{B}$ and $\lambda_\text{T}$ drives $\Langle \Delta \ell_\text{PTR}^2\Rangle$.  When $\theta > \theta_0$, on the other hand, $\Langle \Delta \ell_\text{PTR}^2\Rangle$ is a combination of the point sources' wavelength and angular separations and converges to the sum of the single-wavelength variances in the limit $\theta/\theta_0 \to \infty$.

\section{Conclusions}
In this paper, we studied the effects of anisoplanatism on two-wavelength AO systems.  This analysis stands in contrast to prior works that assumed isoplanatic conditions.  Utilizing Mellin transform techniques, we derived closed-form expressions for the anisoplanatic, two-wavelength PR and PTR OPD variances---two key metrics of AO system performance.  

To validate our work, we compared our theoretical variances to numerical results and found them to be in excellent agreement.  Furthermore, we described the physical significance of our results and introduced the two-wavelength isoplanatic angle---a key predictor of two-wavelength AO performance.  The methods, analysis, and findings presented in this paper will find use in the design and implementation of two-wavelength AO systems for multiple imaging and beam projection applications.

\section*{Open Research}
Version 1.0.1 of \texttt{PLOT\_Two\_Wvl\_PTR\_OPD\_Var\_Anisoplanatism\_Theory.py}, used for evaluating the integrals and Meijer G-functions in the paper and generating Figures~\ref{fig:PR} and~\ref{fig:PTR}, is preserved at doi: 10.5281/zenodo.16568660~\cite{Hyde}.

\acknowledgments
The authors would like to thank the Joint Directed Energy Transition Office for sponsoring this research.  M.W.H.: The views expressed in this paper are those of the author and do not reflect the policy or position of Epsilon C5I, Inc. or Epsilon Systems Solutions, Inc. The U.S. Government is authorized to reproduce and distribute reprints for governmental purposes notwithstanding any copyright notation thereon. Department of Defense Office of Prepublication and Security Review CLEARED For Open Publication 14 Aug 2025.

%
%

\bibliography{refs}

%
%
%
%
%

\end{document}